\documentclass[floats,preprint,prb,aps]{revtex4}

\begin{document}
\preprint{{\em Submitted to Phys. Rev. {\bf E} } \hspace{3.7in}
Web galley}
\title{Superoperator Representation of Nonlinear Response; Unifying Quantum Field and Mode Coupling Theories}
\author{Shaul Mukamel}
\affiliation{Department of Chemistry, University of California,
Irvine, CA 92697-2025
\\}

\date{\today}
\vspace{2.7in}
\begin{abstract}
Computing response functions by following the time evolution of
superoperators in Liouville space (whose vectors are ordinary
Hilbert space operators) offers an attractive alternative to the
diagrammatic perturbative expansion of many-body equilibrium and
nonequilibrium Green functions. The bookkeeping of time ordering
is naturally maintained in real (physical) time, allowing the
formulation of Wick's theorem for superoperators, giving a
factorization of higher order response functions in terms of two
fundamental Green's functions. Backward propagations and the
analytic continuations using artificial times (Keldysh loops and
Matsubara contours) are avoided. A generating functional for
nonlinear response functions unifies quantum field theory and the
classical mode coupling formalism of nonlinear hydrodynamics and
may be used for semiclassical expansions. Classical response
functions are obtained without the explicit computation of
stability matrices.

\vspace{1.0 in} Pacs numbers: $05.30.-d,05.20.Jj,42.65+k,11.10.-z$

\end{abstract}
\maketitle{}

\bigskip

\newpage
\section{Introduction}

An important ingredient in many-body theories is the ability to
factorize averages of products of a large number of operators into
products of averages of pairs. This Wick theorem is common to the
broad arsenal of techniques used for the treatment of quantum and
classical systems alike. Quantized fields are used e.g. in Green
function perturbation theory of many identical bosons or
fermions;~\cite{Negele,Fetter2,Doniach,Abrikosov2,Kadanoff,Mahan,Haug}
Time Dependent Hartree-Fock (TDHF) and Time Dependent Density
Functional (TDDFT) equations of motion of many electron
systems~\cite{Ring} and the Hartree-Fock Bogoliubov equations for
superconductors and Bose Einstein condensates~\cite{Thouless}.
Classical fields are considered in mode coupling theories of
nonlinear hydrodynamics of fluids and
glasses~\cite{Schofield,Ronis}; cumulant $(1/N)$ expansions for
short range interactions in fluids, and Gaussian models of spin
Hamiltonians~\cite{Reichman2,Machta,Denny2,vanZon,Kawasaki,Bouch,Gotze}.

Green function perturbation theory forms the basis for the
powerful Feynman diagrammatic techniques widely used in the
description of many-particle
systems~\cite{Negele,Fetter2,Doniach,Abrikosov2,Kadanoff,Mahan,Haug,Ring,
Thouless}. This formalism is based on expressing quantities of
interest as \emph{time-ordered expansions}. Equilibrium and non
equilibrium Green function
techniques~\cite{Kadanoff,Haug,Economov} employ various types of
contours which, in effect, transform the computation to a time
ordered form in some artificial (unphysical) time variable along
the contour~\cite{Keldysh,Matsubara,KUBO_BOOK}.

The primary goal of this article is to demonstrate that the
description is greatly simplified by employing superoperator
algebra and computing response functions using the density matrix
in Liouville
space~\cite{Mukamel,Chernyaka,Chernyak340,Mukamel358}. One of the
rewards of working in the higher dimensional Liouville space is
that we only need consider time ordered quantities in real
(physical) time and Wick's theorem therefore assumes a
particularly compact form; no special contours or analytic
continuations are necessary. The Hilbert space description
requires a sequence of forward/backward propagations as opposed to
the all-forward representation of response functions in Liouville
space~\cite{48,Sepulveda,7,11,12}. The superoperator approach
provides a unifying framework applicable to quantum and classical
systems, with and without second quantization. It thus connects
field theories with classical mode coupling theories of
fluctuating hydrodynamics. Semiclassical approximations are
developed directly for nonlinear response functions (i.e.,
specific combinations of correlation functions) rather than for
individual correlation functions, which do not have a natural
classical limit and their semiclassical approximations are thus
ill defined. Recent interest in multidimensional Raman techniques
generated considerable activity in modelling multitime correlation
functions~\cite{keyes,Ohmine,Keyes2,Jansen,Tanimura,STRATT,Okumurac,CAO,REICHMAN}.
The mode coupling simulation of correlation functions using
Langevin equations poses many problems~\cite{vanZon,Keyes2}. These
difficulties disappear by modelling the entire response where the
classical limit is uniquely and unambiguously recovered. The
present formalism shows how nonlinear response functions may be
expressed in terms of lower order response of collective
variables~\cite{Chernyak340,Mukamel358,Chernyakb}.

In Section II we discuss two strategies for simulating response
functions. The first, based on the wavefunction in Hilbert space,
does not maintain a full bookkeeping of time ordering whereas the
second, based on the density matrix in Liouville space
does~\cite{Fano,BenReuven,Zwanzig}. A detailed comparison is made
of the physical insight and the numerical effort required in both
pictures. These results form the basis for developing the
many-body Green function perturbation theory in Section III. Using
a generalized superoperator generating functional, we obtain a
time ordered perturbation theory of elementary Liouville space
operators, and derive Wick's theorem for Boson field
superoperators in Section IV. These results are used in Section V
to derive a semiclassical expansion for response functions which
in the classical limit recovers mode coupling theory. The
extension to Fermion fields is made in Section VI and our results
are summarized and discussed in Section VII.

Wick's theorem is based on a perturbative expansion around a
quadratic Hamiltonian and is thus limited to physical situations
when this is a good reference for the actual dynamics. It is given
for Boson fields in Section IV using a closed expression for a
generating functional, and for Fermion fields in Section VI. In
Section V we explore it in coordinate space without using second
quantization. Section II introduces the notation and reviews
previous results. The superoperator algebra of Section III was
used earlier for specific applications (time dependent
Hartree-Fock, fifth Raman
spectroscopy).~\cite{Chernyaka,Chernyak340,Mukamel358,Khidekelc}
This section recasts these earlier results in a more general and
compact notation that sets the stage for the subsequent sections.

\section{Liouville vs. Hilbert Space Description of Quantum
Nonlinear Response}

\underline{Partially Time Ordered, Wavefunction Based Expansion of
Response Functions}

We consider a material system with Hamiltonian $H$, coupled to an
external driving field $E(\tau)$ by the interaction
\begin{equation}\label{2.2t}
 H_{int}(\tau) = - A E(\tau),
\end{equation}
where $A$ is a general dynamical variable. For clarity we assume a
scalar field; Extension to vector fields is straightforward by
introducing tensor notation. The total Hamiltonian $H_T(\tau)$ is
given by
\begin{equation}\label{2.4t}
H_T(\tau) = H + H_{int}(\tau).
\end{equation}

We shall be interested in the expectation value of an operator $B$
of the driven system at time $t$. For a system described by a
wavefunction $\mid\psi_j(t)\rangle$ this is given by $S(t)\equiv
\langle \psi_j (t) | B | \psi_j (t) \rangle$.  A perturbative
calculation of $\mid\psi_j(t)\rangle$ then gives to $n$'th order
in the field
\begin{equation} \label{wave1}
S_j^{(n)}(t) \sum_{m=0}^{n} \langle \psi_j^{(m)} (t) | B |
\psi_j^{(n-m)} (t) \rangle.
\end{equation}
Here $| \psi_{j}^{(m)} \rangle$ denotes the wavefunction to $m$'th
order in $ H_{int}$. If the system is initially in a mixed state
(e.g. Canonical distribution) where state $|j \rangle$ is occupied
with probability $P_j$, we need to average Eq.~(\ref{wave1}) over
that ensemble
\begin{equation} \label{wave2}
S^{(n)} (t) = \sum_j P_j S_j^{(n)} (t).
\end{equation}

Time dependent perturbation theory gives for the linear
response~\cite{Mukamel}
\begin{equation} \label{wave3}
S^{(1)}_j (t) = \frac{i}{\hbar}\int_{- \infty}^{t} d\tau_{1}
\langle \psi_j | U^{\dag} (t-\tau_{1})B U (t-\tau_{1}) A | \psi_j
\rangle E(\tau) + c.c.
\end{equation}
Here $|\psi_{j}\rangle \equiv |\psi_{j}(0) \rangle$ and $U(\tau)$
is the \emph{retarded} evolution operator in Hilbert Space which
propagates the wavefunction forward in time
\begin{equation} \label{B1}
U(\tau) = \theta(\tau)\exp\left(-\frac{i}{\hbar} H\tau\right),
\end{equation}
whereas the \emph{advanced} Green function
\begin{equation} \label{B2}
U^\dagger(\tau) =
\theta(\tau)\exp\left(\frac{i}{\hbar}H\tau\right),
\end{equation}
is responsible for backward propagation. $\theta(\tau)$ denotes
the Heavyside function (0 for $\tau<0$, 1 for $\tau>0$).

For the third order response which describes many of the most
common nonlinear spectroscopies~\cite{Mukamel}, we obtain
\begin{eqnarray} \label{wave4}
&&S^{(3)}_j (t) = \left(\frac{i}{\hbar}\right)^{3} \int_{-
\infty}^t d \tau_1 \int_{-\infty}^t d \tau_2
\int_{-\infty}^{\tau_{2}} d \tau_3 R_a
(t, \tau_3, \tau_2, \tau_1) E (\tau_1) E(\tau_2) E(\tau_3)\\
\nonumber
&& + \left(\frac{i}{\hbar}\right)^{3}\int_{- \infty}^t d \tau_1
\int_{-\infty}^{\tau_{1}} d \tau_2 \int_{-\infty}^{\tau_{2}} d
\tau_3 R_b (t, \tau_3, \tau_2, \tau_1) E (\tau_1) E(\tau_2)
E(\tau_3)+ c.c
\end{eqnarray}
where
\begin{eqnarray} \label{wave5}
R_a (\tau_{4},\tau_{3},\tau_{2},\tau_{1})&=&  \langle \psi_j |
U^\dagger (\tau_{31}) A U^\dagger(\tau_{43}) B U(\tau_{42}) A
U(\tau_{21}) A | \psi_j \rangle\\ \nonumber
R_b (\tau_{4},\tau_{3},\tau_{2},\tau_{1})&=&  \langle \psi_j |
U^\dagger (\tau_{31}) A U^\dagger(\tau_{23}) A
U^\dagger(\tau_{42}) B U(\tau_{41}) A | \psi_j \rangle\\ \nonumber
R_c(\tau_{4},\tau_{3},\tau_{2},\tau_{1}) &=& \langle \psi_j |
U^\dagger (\tau_{41}) B U(\tau_{43}) A U(\tau_{32}) A U(\tau_{21})
A | \psi_j \rangle.
\end{eqnarray}
and we have defined $\tau_{4}\equiv t$ and
$\tau_{ij}\equiv\tau_i-\tau_j$. These equations represent a time
loop of forward and backward propagations~\cite{Schwinger}.
Eq.~(\ref{wave5}) may be alternatively recast using correlation
functions
\begin{eqnarray}
R_{a} (\tau_{4},\tau_{3},\tau_{2},\tau_{1})&=& \langle\psi_{j}|
\hat{A} (\tau_{3}) \hat{B}(\tau_{4})\hat{A}(\tau_{2})\hat{A}
(\tau_{1})|\psi_{j}\rangle\\\nonumber
R_{b}(\tau_{4},\tau_{3},\tau_{2},\tau_{1})&=& \langle\psi_{j}|
\hat{A} (\tau_{3}) \hat{A}(\tau_{2})\hat{B}(\tau_{4})\hat{A}
(\tau_{1})|\psi_{j}\rangle\\\nonumber
R_{c}(\tau_{4},\tau_{3},\tau_{2},\tau_{1})&=& \langle\psi_{j}|
\hat{B} (\tau_{4}) \hat{A}(\tau_{3})\hat{A}(\tau_{2})\hat{A}
(\tau_{1})|\psi_{j}\rangle\\\nonumber
\end{eqnarray}
where we denote operators in the Heisenberg picture by ( $\hat{} $
)

\begin{equation}\label{hat}
\hat{A}(\tau) \equiv U^{\dag}(\tau) A U (\tau)
\end{equation}
The time variables of $R_c$ in Eq.~(\ref{wave4}) are fully time
ordered $(\tau_1\leq\tau_2\leq\tau_3\leq t)$. However, this is not
the case for $R_a$ and $R_b$. By breaking the integrations into
various segments we can maintain full time ordering, and recast
Eq.~(\ref{wave4}) using a response function. This will be done
next through the density matrix expansion.

\underline{Time-Ordered Expansion: Response Functions}

Rather than using a wavefunction, the state of the system can be
described by its density matrix, defined as
\begin{equation} \label{wave6}
\rho (t) = \sum_j | \psi_j (t) \rangle P_j \langle \psi_j (t) |.
\end{equation}
Eqs.~(\ref{wave1}) and (\ref{wave2}) can be alternatively recast
in the form
\begin{equation} \label{wave7}
S^{(n)} (t) = Tr \left[ B \rho^{(n)} (t) \right],
\end{equation}
where
\begin{equation} \label{wave8}
\rho^{(n)} (t) = \sum_{j}\sum_{m = 0}^{n}P_j | \psi_j^{(m)} (t)
\rangle \langle \psi_j^{(n-m)} (t) |
\end{equation}
is the density matrix expanded to the $n$'th order in $H_{int}$.
The expectation value of $B$ to $n$'th order in the field is
obtained by computing the density matrix to n'th order. This
gives~\cite{Mukamel}

\begin{eqnarray}\label{wave9}
&&S^{(n)} (t) = \int_{-\infty}^t d \tau_n \int_{-
\infty}^{\tau_{n}} d \tau_{n-1} \ldots \int_{-\infty}^{\tau_2} d
\tau_1\\ \nonumber
&& R^{(n)} (t, \tau_n, \tau_{n-1}, \ldots, \tau_1)E(\tau_1)E(\tau_2)\ldots E(\tau_n).\\
\nonumber
\end{eqnarray}
Here $R^{(n)}$ is the \emph{n'th order response function}
\begin{equation}\label{3.6t}
R^{(n)} (\tau_{n+1} \ldots \tau_1) = \left( \frac {i}{\hbar}
\right)^n
\nonumber \\
\mathrm{Tr}\left\{[\ldots[[\hat{B}(\tau_{n+1}), \hat{A}(\tau_n)],
\hat{A}(\tau_{n-1})]\ldots, \hat{A}(\tau_1)]\rho_{eq}\right\} .
\end{equation}
which can be alternatively recast as
\begin{equation}
\label{eq2} R^{(n)} (\tau_{n+1} \ldots \tau_1) \! = \!\bigg(
\!\frac {i}{\hbar} \!\bigg)^{\!n} {\rm Tr} \Big\{ \hat{B}(\tau_{n
+ 1}) \Big[ \hat{A}(\tau_n), \dots, [
   \hat{A}(\tau_2), [ \hat{A}(\tau_1), \rho_{eq} ] ] \cdots \Big] \Big\}.
\end{equation}
Note that the time variables $\tau_{j}$ in Eq.~(\ref{wave4}) are
not time-ordered. In contrast, the complete time ordering in
Eq.~(\ref{wave9}) makes the density matrix description most
intuitive and directly connected to experiment~\cite{Mukamel}.

In the density matrix formulation we maintain a simultaneous
bookkeeping of the interactions with the ket and with the bra.
This is why Eq.~(\ref{eq2}) has $2^{n}$ terms, each constituting a
distinct \emph{Liouville space pathway}. The wavefunction
calculation, in contrast, focuses on amplitudes and the various
time orderings of the ket and the bra interactions are lumped
together. Eq.~(\ref{wave1}) has thus only $n+1$ terms. The
different terms in this case simply reflect the order of the
interactions within the bra and within the ket (but not the
relative time ordering of bra and ket interactions!). When the
system interacts with a thermal bath, the $2^n$ terms in
Eq.~(\ref{eq2}) represent very different physical processes and
their separate treatment is absolutely crucial.  The density
matrix separates these terms directly and naturally without the
need for any change of time variables.

The quantum nonlinear response function $R^{(n)}$ is given by a
combination of $(n+1)$ order correlation functions. Response
functions provide a natural link between theory and
experiment~\cite{b}. $R^{(n)}$ is a purely material quantity which
contains all the necessary information for describing n'th order
response. It is independent of the details of a particular
measurement, (e.g. temporal sequences of pulses as well as their
frequencies and wavevectors). The field envelopes enter through
the multitime convolutions in Eq.~(\ref{wave9}). When $S^{(n)}$ is
calculated in terms of the wavefunction without using response
functions (Eq.~(\ref{wave2})) we need to repeat the calculation
for every new realization of the field. $R^{(n)}$ is therefore a
compact and economical way for clarifying the fundamental
relationships among various techniques and their information
content. Since the nonlinear response functions are successively
probing higher order correlation functions, they necessarily carry
additional information as the order n is increased.

\underline{Forward/Backward vs. All-Forward Representation of
Response Functions}

The expression for the response obtained by expanding the density
matrix in powers of the external field (Eq.~\ref{3.6t}), separates
naturally into several contributions, each representing a distinct
time-ordering of the various interactions. The time variables
appearing in Eq.~(\ref{wave9}) are chronologically ordered and
represent successive interactions with the field. In contrast, the
time variables in the wavefunction description are not fully
ordered and consequently have a much less transparent physical
interpretation. $R^{(n)}$ has $2 ^{n}$ terms (Liouville space
pathways) in the density matrix description (Eq.~(\ref{3.6t})) but
only 2n terms using wavefunctions (Eq.~(\ref{wave4})). In practice
we need only compute half of the terms ($2 ^{(n-1)}$ vs. n) since
the other terms are their complex conjugates.

For the linear response Eq.~(\ref{3.6t}) gives
\begin{eqnarray}
\label{wave10060} R^{(1)} (\tau_{2},\tau_1) =
\frac{i}{\hbar}\sum_jP_j\langle\psi_j|U^\dagger
(\tau_{21})BU(\tau_{21})A|\psi_j\rangle + c.c.  .
\end{eqnarray}
The third order response is similarly given by
\begin{equation}
\label{B5}R^{(3)}(\tau_{4},\tau_3,\tau_2,\tau_1)= \left(
\frac{i}{\hbar}\right)^{3} \sum^{4}_{s=1}
R_s^{(3)}(\tau_{4},\tau_3,\tau_2,\tau_1) +c.c. .
\end{equation}
\begin{eqnarray}
\label{wave1000}R^{(3)}_1(\tau_{4},\tau_3,\tau_2,\tau_1)=\sum_jP_j\langle\psi_j|U^\dagger
(\tau_{21})
AU^\dagger(\tau_{32})AU^\dagger(\tau_{43})BU(\tau_{41})A|\psi_j\rangle\\
\nonumber
R^{(3)}_2(\tau_{4},\tau_3,\tau_2,\tau_1)=\sum_jP_j\langle\psi_j|U(\tau_{21})
AU^\dagger(\tau_{31})AU^\dagger(\tau_{43})BU(\tau_{42})A|\psi_j\rangle\\
\nonumber
R^{(3)}_3(\tau_{4},\tau_3,\tau_2,\tau_1)=\sum_jP_j\langle\psi_j|U(\tau_{31})
AU^\dagger(\tau_{21})AU^\dagger(\tau_{42})BU(\tau_{43})A|\psi_j\rangle\\
\nonumber
R^{(3)}_4(\tau_{4},\tau_3,\tau_2,\tau_1)=\sum_jP_j\langle\psi_j|U^\dagger(\tau_{41})
BU(\tau_{43})AU(\tau_{32})AU(\tau_{21})A|\psi_j\rangle
\end{eqnarray}
Unlike Eq.~(\ref{wave4}), Eq.~(\ref{wave9}) allows us to define a
response function since it is fully time ordered.  Note that $R_4
= R_c$, and $R_a+R_b$ correspond to $R_1 + R_2 + R_3$.

Eqs.~(\ref{wave10060}) and ~(\ref{B5}) can be calculated by either
expanding the correlation functions in eigenstates or using
wavepackets in the coordinate representation. Semiclassically it
is possible to expand $\mid\psi_j(t)\rangle$ in coherent states
$\mid\psi_j(t)\rangle = \int\int d{\bf p} d{\bf q} \mid {\bf {p
q}}\rangle \langle {\bf {p q}} \mid\psi_j(t)\rangle$. Each $R_j$
may thus be computed as an average given by a sum over
trajectories moving forward and backward in time as given by the
various $U$ and $U^\dagger$ factors, respectively. Coherent states
provide an over complete basis set~\cite{Glauber}. Powerful
semiclassical approximations were developed for carrying out this
propagation~\cite{48,Sepulveda,7,11,12,148}.

In Eqs.~(\ref{B5}) and~(\ref{wave1000}) we used the density matrix
to derive formal expressions for the response functions, but for
the actual calculation we went back to the wavefunction in Hilbert
space. Since quantum mechanics is usually described in terms of
wavefunctions, wavepacket and semiclassical descriptions are
normally developed for wavefunctions. It is possible however to
construct an alternative forward propagating wavepacket picture by
staying with the density matrix in Liouville space all the way. To
that end we represent the time dependent density matrix as
\begin{equation} \label{Buba}
\rho(t)=U(t)\rho(0)U^\dagger(t)\equiv {\cal G}(t)\rho(0).
\end{equation}
The first equality is the common representation where we treat
$\rho(t)$ as an \emph{operator in Hilbert space}.  In the second
equation we consider $\rho(t)$ as a \emph{vector in Liouville
space}. We further introduce the Liouville Space evolution
operator
\begin{eqnarray}\label{B}
{\cal G}(t)=\theta(t)\exp\left(-\frac{i}{\hbar}Lt\right),
\end{eqnarray}
where $ LA\equiv[H,A],$ is the Liouville operator.

We shall denote superoperators by a subscript $\nu=L,R$ where the
operators $ A_L$ and $ A_R$ act on the ket (left) and bra (right)
of the density matrix ($A_L B \equiv AB$ and $A_R B \equiv
BA$)~\cite{chernyak}. We further define the equilibrium
distribution function
\begin{equation}\label{hungry}
\rho_{eq}=\sum_jP_j|\psi_j(0)\rangle\langle \psi_j(0)|.
\end{equation}
Adopting this notation for Eq.~(\ref{eq2}) yields for the linear
response
\begin{equation}
\label{B4} R^{(1)} (\tau_2,\tau_1) = \mathrm{Tr}\left[B_L{\cal
G}(\tau_{21})A_L\rho_{eq}\right] +c.c.
\end{equation}

and for the third order response
\begin{eqnarray}
\label{B6}R^{(3)}_1(\tau_4,\tau_3,\tau_2,\tau_1)&=& \mathrm{Tr}
[B_L {\cal G}(\tau_{43})A_R {\cal G}(\tau_{32})A_R {\cal
G}(\tau_{21})A_L{\rho_{eq}}]\\\nonumber
R^{(3)}_2(\tau_4,\tau_3,\tau_2,\tau_1)&=& \mathrm{Tr} [B_L {\cal
G}(\tau_{43})A_R {\cal G}(\tau_{32})A_L {\cal
G}(\tau_{21})A_R{\rho_{eq}}]\\\nonumber
R^{(3)}_3(\tau_4,\tau_3,\tau_2,\tau_1)&=& \mathrm{Tr} [B_L {\cal
G}(\tau_{43})A_L {\cal G}(\tau_{32})A_R {\cal
G}(\tau_{21})A_R{\rho_{eq}}]\\\nonumber
R^{(3)}_4(\tau_4,\tau_3,\tau_2,\tau_1)&=& \mathrm{Tr} [B_L {\cal
G}(\tau_{43})A_L {\cal G}(\tau_{32})A_L {\cal
G}(\tau_{21})A_L{\rho_{eq}}].\\\nonumber
\end{eqnarray}
\emph{Note that since the density matrix needs only to be
propagated forward, Eqs.~(\ref{B6}) only contain the forward
propagator ${\cal G}(t)$ and not its Hermitian conjugate ${\cal
G}^\dagger(t)$, which describes backward propagation.} This is in
contrast with the Hilbert space expression (Eq.~(\ref{wave1000}))
which contains both $U(\tau)$ and $U^{\dag}(\tau)$.

Similar to the wavefunction picture, the response functions may be
computed by sums over states or by semiclassical wavepackets
\begin{equation}
\label{B7}\rho_j^{(n)}(t)=\int\int d{\bf p}d{\bf q} d{\bf p}'
d{\bf q}'\mid {\bf p}'{\bf q}'\rangle\langle {\bf p}'{\bf q}'\mid
\rho^{(n)}_j(t)\mid {\bf p}{\bf q} \rangle\langle {\bf p}{\bf
q}\mid.
\end{equation}

Each term {\em (Liouville space path)} in Eq.~(\ref{B6}) can be
recast in the form~\cite{Mukamel,Yan}
\begin{equation}
R_{j}^{(3)}(\tau_{4},\tau_{3},\tau_{2},\tau_{1})= \mathrm{Tr}
[B_{L}\rho_{j}^{(n)}(t)]
\end{equation}
where $\rho_{j}^{(n)}(t)$ is a density matrix generating function
for path $j$ which can be computed using two forward moving
trajectories representing the simultaneous evolution of the ket
and the bra~\cite{Sepulveda,Khidekelc}. In the wavefunction
representation we act on the ket only. Propagating the bra forward
is equivalent to propagating the ket backward. By keeping track of
both bra and ket simultaneously we can enjoy the physically
appealing all-forward evolution. Since the various Liouville space
pathways are complex quantities, they interfere when added. This
interference may result in dramatic effects.

A systematic approach for computing the response functions will be
developed in the next section.

\section{Superoperator Algebra and the Time Ordered
Perturbative Expansion of Response Functions}

In Eqs.~(\ref{B4}) and~(\ref{B6}) we introduced the indices $L$
and $R$ to denote the action of a superoperator from the left or
the right. In the following manipulations, in particular for the
sake of developing a semiclassical picture, it will be useful to
define their symmetric $(\nu=+)$ and antisymmetric $(\nu=-)$
combinations~\cite{Chernyaka}
\begin{equation}\label{A3t}
A_- \equiv A_L - A_R; \hspace{0.2in} A_+ \equiv \frac {1} {2} (A_L
+ A_R).
\end{equation}
Recasting these definitions in Hilbert space using ordinary
operators we get $A_{+}X \equiv \frac{1}{2}(AX + XA) $ ; $A_{-} X
\equiv AX -XA $, $X$ being an arbitrary operator. Hereafter we
shall use Greek indices to denote superoperators $A_\nu$ with
either $\nu=L,R$ or $\nu=+,-$.

Hereafter we shall consider operators that depend parametrically
on time. This time dependence can be either in the Heisenberg
picture $\hat{A}_{\nu}(\tau)$ (Eq.~(\ref{32})) or in the
interaction picture $\tilde{A}_{\nu}(\tau)$ (Eq.~(\ref{7})). By
introducing a \emph{time ordering operator} $T$ \emph{for
superoperators in Liouville space}, we can freely commute various
operators without worrying about commutations. $T$ takes any
product of superoperators and reorders them in ascending times
from right to left. More precisely we define
\begin{equation}
\label{3} T A_{\nu}(\tau_{1})B_{\mu}(\tau_{2}) =
\left\{\begin{array}{ll}
 A_{\nu}(\tau_{1})B_{\mu}(\tau_{2})& \tau_{2}<\tau_{1}\\
 B_{\mu}(\tau_{2})A_{\nu}(\tau_{1})& \tau_{1}<\tau_{2}\\
 \frac{1}{2}[A_{\nu} (\tau_1) B_{\mu} (\tau_1) + B_{\mu}(\tau_1)
A_{\nu}(\tau_1)]&\tau_{2}=\tau_{1}\end{array}\right.
\end{equation}
where $A_{\nu}(\tau)$ is either $\hat{A}_{\nu}(\tau)$ or
$\tilde{A}_{\nu}(\tau)$. $T$ orders all superoperators such that
time decreases from left to right: The latest operator appears in
the far left and so forth. This is the natural time ordering which
follows chronologically the various interactions with the density
matrix~\cite{Ohmine}. The precise order in which superoperators
appear next to a $T$ operator is immaterial since at the end the
order will be fixed anyhow by $T$. For example, $T$ before an
exponent means that each term in the Taylor expansion of this
exponent should be time-ordered.

We next introduce the Heisenberg picture for superoperators whose
time evolution is governed by the Liouville operator
\begin{equation}
\label{32} \hat{A}_{\nu}(\tau) \equiv
\hat{\mathcal{G}}^{\dag}(\tau,0)A_{\nu} \hat{\mathcal{G}}(\tau,0)
\end{equation}
with
\begin{equation}
\hat{\mathcal{G}}(\tau_{2},\tau_{1}) = \theta (\tau_{2}-
\tau_{1})\exp \,\left[-\frac{i}{\hbar} L(\tau_{2}-
\tau_{1})\right].
\end{equation}
Eq.~(\ref{32}) is the Liouville space analogue of Eq.~(\ref{hat}).
The expectation value of $B$
\begin{equation}
\label{30}
 S (t) = \mathop{\rm Tr} [ B \rho (t)],
\end{equation}
may now be represented in a form
\begin{equation}\label{3.3t}
  S(t) = \langle T\hat{B}_+(t) \exp \left [\frac {i} {\hbar} \int_{-\infty}^{t} d\tau
  E(\tau)\hat{A}_{-}(\tau) \right] \rangle.
\end{equation}
The operator $\hat{B}_+(t)$ corresponds to the observation time,
whereas $\hat{A}_-(\tau_j)$ represent various interactions with
the external field at time $\tau$, and $\langle \cdots \rangle$
denotes averaging with respect to the equilibrium density matrix
$\rho_{eq}$.
\begin{equation}
\langle F \rangle \equiv \mathrm{Tr}[F\rho_{eq}]
\end{equation}

By expanding the exponent in the r.h.s. of Eq.~(\ref{3.3t}) in
powers of $E(\tau)$, we obtain for the response functions.

\begin{equation}\label{3.5t}
  R^{(n)} (\tau_{n+1}\ldots \tau_1) \equiv
  \left ( \frac {i} {\hbar} \right )^n
  \langle \hat{B}_+ (\tau_{n+1}) \hat{A}_- (\tau_n) \ldots \hat{A}_-(\tau_1) \rangle.
\end{equation}
Eq.~(\ref{3.5t}) is merely a compact notation for Eq.~(\ref{eq2}).
It should be emphasized that all time arguments are fully ordered
$\tau_1 \leq \tau_2 \ldots \leq \tau_{n+1}$. The Liouville space
correlation function in the r.h.s. represents a combination of
ordinary (Hilbert space) correlation functions.

Eq.~(\ref{3.5t}) may be evaluated directly only for simple models.
To convert it into a general computational tool we need to develop
a perturbation theory for response functions based on time ordered
superoperators. To that end we partition the Hamiltonian into a
simple solvable, (usually quadratic) part $H_{0}$ and a
perturbation $V$
\begin{equation}
\label{2}
 H = H_{0} + V,
\end{equation}
and introduce the Heisenberg and interaction pictures. We define
the Liouville operators $ L = L_{0} + V_{-}$ corresponding to
Eq.~(\ref{2}) where $L_0 \equiv (H_0)_{-}$ i.e., $L_0 X \equiv H_0
X -XH_0$. The time evolution operator with respect to $L_0$ is
\begin{equation}
\label{4} \mathcal{G}_{0}(\tau_{2},\tau_{1})=\theta (\tau_{2}-
\tau_{1})\exp\left[-\frac{i}{\hbar}L_{0}(\tau_{2}-\tau_{1})\right].
\end{equation}
The total (Heisenberg) time evolution operator with respect to $L$
will be denoted $\hat\mathcal{G} (\tau_2, \tau_1)$. We can then
write
\begin{eqnarray}
\label{5} \hat\mathcal{G} (\tau_{2},\tau_{1}) &=&
\mathcal{G}_{0}(\tau_{2},\tau_{1})\tilde\mathcal{G}(\tau_{2},\tau_{1})
\end{eqnarray}
where $\tilde\mathcal{G}$ is the time evolution operator in the
\emph{interaction picture}
\begin{eqnarray}
\label{6} \tilde\mathcal{G}(\tau_{2},\tau_{1})&=& T \exp
\left[-\frac{i}{\hbar} \int_{\tau_{1}}^{\tau_{2}} d\tau_{2}
\tilde{V}_{-}(\tau)\right].
\end{eqnarray}

Throughout this paper we use a hat ( $\hat{} $ ) to denote
operators in the \emph{Heisenberg picture} (Eq.~(\ref{32})) and a
tilde ( $\tilde{} $ ) for operators in the interaction picture,
i.e.
\begin{eqnarray}
\label{7}
 \tilde{A}_\nu (\tau)\equiv \mathcal{G}_{0}^{\dag} (\tau, 0)
A_{\nu} \mathcal{G}_0(\tau, 0)&\\\nonumber &\nu = +,-,\,\, or \,\,
\nu = L,R.
\end{eqnarray}

The equilibrium density matrix of the interacting system can be
generated from the density matrix of the noninteracting system
$(\rho_0)$ by an adiabatic switching of the coupling $V$,
resulting in
\begin{equation}
\label{Interaction} \rho_{eq} = \tilde{\mathcal{G}} (0, -\infty)
\rho_{0}.
\end{equation}

For isolated system at zero temperature, Eq.~(\ref{Interaction})
generates the ground state density matrix of the interacting
system, starting with the noninteracting ground state. This is the
procedure of Gell-mann and Low~\cite{Gellman}. At zero temperature
the zero order ground state evolves into the actual normalized
ground state and hence Eq.~(\ref{Interaction}) need not have a
denominator. Note that in the wavefunction (Gell-mann-Low)
formulation of adiabatic switching, the wavefunction acquires a
singular phase which must be cancelled by a denominator given by
the closed loop $S$ matrix; the Liouville space expression is
simpler since the phase never shows up. A remarkable point is that
Eq.~(\ref{Interaction}) holds as well at finite temperatures
provided the system is coupled to a bath at constant temperature.
This is a thermodynamic adiabatic switching where the populations
of adiabatic states change and equilibrate with the bath at all
times~\cite{Jarzynski,Frankel,MukJarz}. It is distinct from the
adiabatic switching of an isolated quantum system where the
populations of adiabatic states do not change~\cite{Wilczek}.

At finite temperatures we start with a grand canonical
distribution
\begin{equation}
\label{12}
 \rho_{0} = \frac{\exp[ -\beta(H_{0}-\mu N)]}{Tr \exp
[-\beta (H_{0}-\mu N)]}
\end{equation}
Where $\beta = (k_{B}T)^{-1}$ ($k_{B}$ is a Boltzmann constant);
$\mu$ is a chemical potential, $N$ is the number operator of
particles, and Eq.~(\ref{Interaction}) generates the distribution.

\begin{equation}
\label{12a}
 \rho_{eq} = \frac{\exp[ -\beta(H-\mu N)]}{Tr \exp
[-\beta (H-\mu N)]}
\end{equation}
We now have all the ingredients required for computing the
response. Let us start with the linear response function
\begin{eqnarray}
R^{(1)} (\tau_{2},\tau_{1})&=& \frac{i}{\hbar} \langle
{\hat{B}_{+}}(\tau_2) {\hat{A}_{-}} (\tau_1)\rangle.
\end{eqnarray}
Using Eqs.~(\ref{32}) and~(\ref{5}) we obtain
\begin{eqnarray}
R^{(1)} (\tau_{2},\tau_{1})&=& \frac{i}{\hbar}  Tr
[\tilde{\mathcal{G}}^{\dag}(\tau_{2}, 0) \tilde{B}_{+}(\tau_{2})
\tilde\mathcal{G}(\tau_{2}, 0)
\tilde\mathcal{G}^{\dag}(\tau_{1},0)
\tilde{A}_{-}(\tau_{1})\tilde\mathcal{G} (\tau_{1},
0)\tilde\mathcal{G} (0,-\infty)\rho_{0}]
\end{eqnarray}
The last $\tilde\mathcal{G}^{\dag}(\tau_{2}, 0)$ can be neglected
since it does not affect the trace. Also
\begin{equation}
\tilde{\mathcal{G}}(\tau_{2},
0)\tilde{\mathcal{G}}^{\dag}(\tau_{1},
0)=\tilde{\mathcal{G}}(\tau_{2}, \tau_{1}),
\end{equation}
which gives
\begin{eqnarray}
\label{14} R^{(1)}(\tau_2, \tau_1)&=& \frac{i}{\hbar}\mathrm{Tr}
[\tilde{B}_{+}(\tau_{2})\tilde{\mathcal{G}}(\tau_{2},\tau_{1})\tilde{A}_{-}
(\tau_{1})\tilde{\mathcal{G}}(\tau_{1},-\infty)\rho_{0}].
\end{eqnarray}

The time ordering operator allows us to express Eq.~(\ref{14}) in
the compact form
\begin{eqnarray}
\label{49} R^{(1)} (\tau_2, \tau_1)&=& \frac{i}{\hbar}
\left\langle T \tilde{B}_{+}(\tau_{2})\tilde{A}_{-}(\tau_{1})\exp
\left[-\frac{i}{\hbar} \int_{-\infty}^{\tau_{2}} d\tau
\tilde{V}_{-}(\tau)\right]\right\rangle_0
\end{eqnarray}
where we define averaging with respect to the density matrix
$\rho_{0}$ of the noninteracting system
\begin{equation}
\langle F \rangle_{0}\equiv \mathrm{Tr}[F\rho_{0}].
\end{equation}
Eq.~(\ref{49}) can be immediately generalized for the response to
arbitrary order
\begin{eqnarray}\label{3.33t}
  S(t) = \langle T \tilde{B}_+(t)\exp \,\,\left[-\frac{i}{\hbar}
\int_{-\infty}^{t} d\tau \tilde{V}_{-}(\tau)\right] \exp \left
[\frac {i} {\hbar} \int_{-\infty}^{t} d\tau
 E(\tau)\tilde{A}_{-}(\tau) \right] \rangle_0.
\end{eqnarray}

Expanding Eq.~(\ref{3.33t}) to n'th order in the external field
gives
\begin{eqnarray}
\label{non_resp} R^{(n)}(\tau_{n+1} \cdots \tau_{1}) &=&
 \left( \frac {i} {\hbar} \right )^n \langle T
\tilde{B}_+(\tau_{n+1}) \tilde{A}_-(\tau_{n})\cdots
\tilde{A}_-(\tau_{1}) \exp \,\,[ - \frac{i}{\hbar}
\int_{-\infty}^{\tau_{n+1}} d\tau \tilde{V}_{-}(\tau)]\rangle_0,
\end{eqnarray}
where we recall that
\begin{eqnarray}
\tilde{X}(\tau) = \exp \left(\frac{i}{\hbar}L_{0}\tau\right) X\;
\exp \left(-\frac{i}{\hbar}L_{0}\tau\right)&&\\\nonumber && X=A_+,
A_-, V_{-}
\end{eqnarray}
The Taylor expansion of the exponent in the r.h.s of
Eq.~(\ref{non_resp}) finally gives
\begin{eqnarray}\label{non_resp_exp}
  R^{(n)}(\tau_{n+1}, {\ldots} \tau_1) & = & \sum_{m=0}^\infty
   \frac{(-1)^m}{m!} \left(\frac {i} {\hbar} \right )^{m+n} \int_{-\infty}^{\tau_{n+1}}
   d\tau_1' {\dots} \int_{-\infty}^{\tau_{n+1}} d\tau'_m \nonumber \\
    & \times  & \langle T \tilde{B}_+(\tau_{n+1}) \tilde{A}_- (\tau_n) \ldots
\tilde{A}_-
    (\tau_1) \tilde{V}_- (\tau_m')
   \ldots \tilde{V}_-(\tau'_1) \rangle_0.
\end{eqnarray}

Eq.~(\ref{non_resp_exp}) constitutes the interaction-picture
representation of the correlation function Eq.~(\ref{3.5t})
~\cite{Chernyaka,Chernyakb}  All superoperators in this expression
should be time ordered chronologically  from right (early time) to
left (late time). This forms the basis for formulating a field
theory and Wick theorem in Liouville space in the next section.

Note that simple time ordering in Liouville space is a more
complex operation when recast in Hilbert space. This is the
essence of why using superoperators makes the bookkeeping
straightforward. To demonstrate that, let us take $R^{(2)}$ (we
use the Heisenberg picture but the argument holds as well in the
interaction picture,  where we  should  simply  replace  all\\ (
$\hat{}$ ) by ( $\tilde{}$ )).
\begin{equation}
R^{(2)}(\tau_3,\tau_2,\tau_1)=\left(\frac{i}{\hbar}\right)^{2}\langle
T\hat{B}_{+}(\tau_3)\hat{A}_{-}(\tau_2)\hat{A}_{-}(\tau_1)\rangle
_{0} .
\end{equation}
We need to apply the superoperators in a time ordered fashion (in
Liouville space) i.e., first $\hat{A}_{-}(\tau_1)$, then
$\hat{A}_{-}(\tau_2)$ and finally $\hat{B}_{+}(\tau_3)$.
Separating all possible actions for the left and the right we get
\begin{eqnarray}
\label{58} 2 \,\mathrm{Tr}[ T
\hat{B}_{+}(\tau_3)\hat{A}_{-}(\tau_2)\hat{A}_{-}(\tau_1)
\rho_{eq}] =
\\\nonumber \mathrm{Tr} [\hat{B}(\tau_3) \hat{A}(\tau_2)
\hat{A}(\tau_1)\rho_{eq}]&+& \mathrm{Tr} [\hat{A}(\tau_2)
\hat{A}(\tau_1)\rho_{eq}\hat{B}(\tau_3)]\\\nonumber
-\mathrm{Tr}[\hat{B}(\tau_3)\hat{A}(\tau_2)\rho_{eq}\hat{A}
(\tau_1)]&-& \mathrm{Tr}
[\hat{A}(\tau_2)\rho_{eq}\hat{A}(\tau_1)\hat{B}(\tau_3)]\\\nonumber
-\mathrm{Tr} [\hat{B}(\tau_3)\hat{A}(\tau_1)\rho_{eq}\hat{A}
(\tau_2)]&-& \mathrm{Tr}[\hat{A}(\tau_1)\rho_{eq}\hat{A}
(\tau_2)\hat{B}(\tau_3)]\\\nonumber + \mathrm{Tr}
[\hat{B}(\tau_3)\rho_{eq} \hat{A}(\tau_1)\hat{A} (\tau_2)] &+&
\mathrm{Tr} [\rho_{eq} \hat{A}(\tau_1)\hat{A}
(\tau_2)\hat{B}(\tau_3)]
\end{eqnarray}

In Hilbert space (r.h.s of Eq.~(\ref{58})) all operators which act
on $\rho_{eq}$ from the left are time ordered and the time
increases as we go to the left starting with $\rho_{eq}$. All
right operators are ordered in the opposite way: Time increases as
we go to the right starting with $\rho_{eq}$. This mixture of
positive and negative time ordering coming from the evolution of
the ket (left) and the bra (right), respectively is what
complicates the bookkeeping of ordinary operators in Hilbert
space. This is in marked contrast with Liouville space (l.h.s. of
Eq.~(\ref{58})) where we keep track of the left and right labels
of the various interactions. Consequently all superoperators are
always positively time ordered in real, physical time which makes
the formulation of a Wick theorem possible.

\section{The Cumulant Expansion and Wick's Theorem for Boson Superoperators}

So far we considered four types of operators which enter
Eq.~(\ref{non_resp}): the reference Hamiltonian $H_0-\mu N$; $A$,
representing the coupling to the external field; $V$, representing
the part of the Hamiltonian to be treated perturbatively, and the
desired observable $B$. To proceed further we need to introduce
the concept of \emph{elementary operators}. Any dynamical system
can ultimately be described by a basic set of operators whose
commutators (or anticommutators) are $c$ numbers. Examples for
elementary operators with commuting algebra (ECA) are the
canonical variables $[Q_\alpha, P_\beta] =
i\hbar\delta_{\alpha\beta}$ and Boson operators, $[a_\alpha,
a_\beta^{\dag}]=\delta_{\alpha\beta}$ used to describe systems of
identical Bosons in second quantization. Second quantized fermions
are described by elementary operators with anticommuting algebra
(EAA) $\{c_\alpha, c_\beta^{\dag}\}= \delta_{\alpha\beta}$. The
operators $X=A,B,V,H_{0}$ and $N$ are some functions of these
elementary operators.

We choose  our reference to be a quadratic Hamiltonian given by a
bilinear combination of elementary field operators.
\begin{equation}
H_{0}=\int dx T(x)\psi^{\dag}(x)\psi(x)
\end{equation}
or using creation/annihilation operators
\begin{equation}
H_{0}= \sum_{r,s}T_{rs} a_{r}^{\dag}a_{s}
\end{equation}
where
\begin{equation}
\psi (x) =\sum_{s}\varphi_{s}(x)a_{s}
\end{equation}
and $\varphi_{s}$ is a single particle basis set. For Bosons,
these operators satisfy the commutation relations
\begin{equation}
\label{60} [a_{s}, a_{r}^{\dag}]=\delta_{rs}
\end{equation}
and
\begin{equation}
[\psi(x),\psi^{\dag}(x')]=\delta (x-x').
\end{equation}
For Fermions, Eq.~(\ref{60}) should be replaced by an
anticommutator. Our elementary set of operators is thus the set
$a_{s}, a_{s}^{\dag}$ or the field operators
$\psi(x),\psi^{\dag}(x)$. The following arguments hold for
Fermions as well, however the derivation is simpler for Boson
fields with ECA. We shall therefore focus on Bosons first, and the
extension to Fermion fields will be presented in Section VI.

We will denote the elementary operators as $Q_j$ and introduce the
corresponding superoperators $Q_{j\nu}$ $\nu=L, R, +, -$. We first
note that the superoperator corresponding to any function of
$Q_{j}$ can be expressed in terms of $Q_{j+}$ and $Q_{j-}$, i.e.,
\begin{eqnarray}
\label{63} [f(Q_{j})]_{-} \equiv f (Q_{jL}) - f (Q_{jR}) &=&
f\left(Q_{j+}+ \frac{1}{2} Q_{j-}\right) -
f\left(Q_{j+}-\frac{1}{2} Q_{j-}\right),
\end{eqnarray}

and
\begin{eqnarray}
\label{63b} 2\,[f(Q_{j})]_{+} \equiv f (Q_{jL}) + f (Q_{jR}) &=&
f\left(Q_{j+}+ \frac{1}{2} Q_{j-}\right) +
f\left(Q_{j+}-\frac{1}{2} Q_{j-}\right).
\end{eqnarray}
For example
\begin{equation}
(Q_{j}^{2})_{+} = Q_{j+}^{2}+ \frac{1}{4} Q_{j-}^{2} ,
\end{equation}
and
\begin{equation}
(Q_{j}^{2})_{-} = Q_{j+} Q_{j-} + Q_{j-} Q_{j+} .
\end{equation}
Using these rules (and additional useful relations given in
Appendix A) we can expand $B_{+}(\tau), A_{-}(\tau)$ and
$V_{-}(\tau)$ in a Taylor series in $Q_{j+}$ and $Q_{j-}$,
converting the time ordered product of superoperators in
Eq.~(\ref{non_resp_exp}) into a time ordered product of elementary
operators. We thus need to calculate
\begin{equation}
\label{63a} W \{j_{m}\nu_{m}\tau_{m}\} \equiv \langle T
\tilde{Q}_{j_N \nu_N}(\tau_N)\ldots
\tilde{Q}_{j_1\nu_1}(\tau_1)\rangle{}_0,
\end{equation}
where $\nu_{1,\ldots,}$ $\nu_N=\pm$ and $j_{m}$ runs over the
various operators. The number $N$ of operators in such products
that enter the computation of $R^{(n)}$ is greater than $n+1, N
\geq n+1$. The reasons are (i) $A_{\nu}, B_{\nu }$ may be
nonlinear functions of elementary operators and we use
Eq.~(\ref{63}) and the formulas of Appendix A to express them as
products of $Q_{\nu}$. (ii) The expansion in $V_{-}$ adds more
operators to the product.

To compute $W$ we define a \emph{superoperator generating
functional}
\begin{eqnarray}
S (\{J (t)\})= \left\langle T\exp\left[\sum_{j\nu}\int J_{j
\nu}(\tau) \tilde{Q} _{j \nu} (\tau) d\tau\right]\right\rangle_{0}
\end{eqnarray}
Time ordered correlation functions of superoperators can be
obtained from the generating functional by functional derivatives
\begin{eqnarray}
\label{68}
 W \{j_{m}\nu_{m}\tau_{m}\}=\frac{\partial}{\partial
J_{j_{1}\nu_{1}}(\tau_{1})}\ldots \frac{\partial}{\partial
J_{j_{N}\nu_{ N}}(\tau _{N})} S \{J(t)\} \bigg|_{J=0}
\end{eqnarray}

Since the Hamiltonian is quadratic, the generating functional may
be computed exactly using the second order cumulant expansion.
This gives
\begin{eqnarray}
\label{65} && S (\{J (t)\})= \exp \left\{
\sum_{j,k}\int_{-\infty}^{\infty}d \tau_{2}
\int_{-\infty}^{\tau_{2}}d \tau_{1}  \right. \\\nonumber &&
[-i\hbar J _{j+}(\tau_{2}) J _{k-}(\tau_{1})
G_{jk}^{+-}(\tau_{2}-\tau_{1}) + J _{j+}(\tau_{2}) J
_{k+}(\tau_{1}) G_{jk}^{++}(\tau_{2}-\tau_{1})] \Bigg\},
\end{eqnarray}
where we have introduced the two fundamental Liouville space Green
functions.
\begin{eqnarray}
\label{66}
 G_{jk}^{+-}(\tau_{2}-\tau_{1})= \frac{i}{\hbar}\langle
T\tilde{Q}_{j+}(\tau_{2}) \tilde{Q}_{k-}
(\tau_{1})\rangle_{0}\\\nonumber G_{jk}^{++}(\tau_{2}-\tau_{1})=
\langle T\tilde{Q}_{j+}(\tau_{2}) \tilde{Q}_{k+}
(\tau_{1})\rangle_{0}.
\end{eqnarray}
Using Eq.~(\ref{A3t}) we can recast these Green functions in
Hilbert space
\begin{eqnarray}
\label{67}
 G_{jk}^{+-}(\tau) &=& \frac{i}{\hbar}\theta (\tau)\left[\langle
\tilde{Q}_{j}(\tau)\tilde{Q}_{k}(0)\rangle_{0}-\langle
\tilde{Q}_{j}(0)\tilde{Q}_{k}(\tau)\rangle_{0}\right]
\end{eqnarray}
\begin{eqnarray}
\label{67a} G_{jk}^{++}(\tau) &=& \frac{1}{2}[\langle
\tilde{Q}_{j}(\tau)\tilde{Q}_{k}(0)\rangle_{0}+\langle
\tilde{Q}_{k}(0)\tilde{Q}_{j}(\tau)\rangle_{0}].
\end{eqnarray}
The $\hbar^{-1} $ factor in $G^{+-}$ was introduced for making the
classical limit more transparent (see next section); since with
this factor $G^{+-}$ has a well defined classical limit. Note that
since the trace of a commutator vanishes identically, in a time
ordered product the superoperator to the far left must be a
``$+$''. The Green functions $G^{--}$ and $G^{-+}$ thus vanish
identically and we only have two fundamental Green functions
$G^{++}$ and $G^{+-}$. Note also that $G^{++} (\tau)$ is finite
for positive and negative $\tau$ whereas $G^{+-} (\tau)$ vanishes
for $\tau < 0$. Eq~(\ref{65}) is an extremely compact expression
which can readily be used to compute response functions to
arbitrary order.

The two fundamental Green functions can be expressed in terms of
the matrix of spectral densities $C_{ij}(\omega)$ defined as the
Fourier transform of
$G^{+-}$~\cite{Mukamel,Chernyakb,43,Redfield}:
\begin{eqnarray}\label{4.8t}
G_{ij}^{+-}(\tau) & = & 2\theta(\tau)
 \int_{-\infty}^\infty \frac {d\omega} {2\pi} C_{ij} (\omega) \sin
 (\omega\tau).
\end{eqnarray}
We then have
\begin{eqnarray}\label{4.8s}
 G_{ij}^{++}(\tau)  & = & \hbar \int_{-\infty}^\infty \frac
 {d\omega} {2\pi} C_{ij} (\omega) \coth \left(\frac{\hbar\omega}{2k_{B}T} \right)
 \cos \,(\omega\tau).
\end{eqnarray}
The Wick theorem for superoperators then follows from
Eqs.~(\ref{68}) and ~(\ref{65}) and can now be stated as follows:
\begin{eqnarray}
\label{69} && \langle T \tilde{Q}_{j_{1}\nu_{1}}(\tau_{1})\ldots
\tilde{Q}_{j_{N}\nu_{N}}(\tau_{N})\rangle_{0} =\\\nonumber &&
\sum_{p} \langle T \tilde{Q}_{j_{a}\nu_{a}}(\tau_{a})
\tilde{Q}_{j_{b}\nu_{b}}(\tau_{b})\rangle_{0} \ldots \langle T
\tilde{Q}_{j_{p}\nu_{p}}(\tau_{p})
\tilde{Q}_{j_{q}\nu_{q}}(\tau_{q})\rangle_{0}
\end{eqnarray}
Here $j_{a}\nu_{a}\ldots j_{q}\nu_{q}$ is a permutation of
$j_{1}\nu_{1}\ldots j_{N}\nu_{N}$ and the sum runs over all
possible permutations, keeping the time ordering. Since only
$G^{++}$ and $G^{+-}$ survive many of the terms will vanish.

Wick's theorem makes it possible to develop Feynman diagram
perturbative techniques which express the linear and non-linear
responses of the interacting system in terms of the two
fundamental Green functions. This theorem is useful whenever a
quadratic reference is adequate and non quadratic parts of the
Hamiltonian can be treated perturbatively. It states that
multipoint correlation functions of systems with quadratic Boson
Hamiltonians may be factorized into products of two-point
correlation functions of the primary coordinates.

\section{Mode Coupling and Semiclassical Response of Boson Fields}

Eq.~(\ref{eq2}) contains $2^n$ terms representing all possible
``left'' and ``right'' actions of the various commutators. Each
term corresponds to a Liouville-space path and can be represented
by a double-sided Feynman diagram~\cite{Mukamel}. The various
correlation functions interfere and this gives rise to many
interesting effects such as new resonances. The $(i/\hbar)^n$
factor indicates that individual correlation functions do not have
an obvious classical limit. The entire response function must,
however, have a classical limit. When the various correlation
functions are combined, the $(i/\hbar)^n$ factor is cancelled as
$\hbar$ tends to $0$, and one obtains the classical response,
independent of $\hbar$. The elimination of $\hbar$ for higher
nonlinearities requires a delicate interference among all $2^n$
correlation functions.

The terms contributing to $R^{(n)}$ (Eq.~(\ref{non_resp_exp}))
will generally have a $(i/\hbar)^{n+p}$ factor where $p$ is the
order in $V_{-}$. This factor must be cancelled as $\hbar
\rightarrow 0$ to ensure a well defined classical limit. This is
guaranteed since by Wick's theorem we will have $n+p$ $G^{+-}$
terms, each carrying an $\hbar$ factor. In the classical limit we
set $\coth (\hbar\omega/2 k_{B}T)\cong 2 k_{B}T/\hbar\omega$. We
then see from (Eq.~\ref{4.8t}) that the two Green functions are
simply connected by the classical fluctuation relation.
\begin{equation}
\label{73A} G^{+-}(\tau)=-\theta (\tau)
\frac{1}{k_{B}T}\frac{d}{d\tau} G^{++}(\tau).
\end{equation}
$G^{+-}$ is independent on $\hbar$. The factor $\hbar\coth
(\hbar\omega/2 k_{B}T)=\hbar/ \tanh (\hbar\omega/2 k_{B}T)$ is
analytic in $\hbar$ and can be expanded in a Taylor series, thus
yielding a semiclassical expansion of the response. To obtain the
classical limit we need to keep $\hbar$ in the generating
functional, perform the $\hbar$ expansion (since response
functions are generally analytic in $\hbar$) and only then send
$\hbar \rightarrow 0$. Setting this limit at the right step is
essential for developing a proper semiclassical expansion.
Classical response functions may not be computed using classical
trajectories alone: The response depends on the vicinity of a
trajectory. One needs to run a few closely lying trajectories that
interfere. Formally this can be recast using stability matrices
which carry the information on how a perturbation of a trajectory
at a given time effects it at a later time. The repeated
computation of the stability matrix greatly complicates purely
classical simulations~\cite{Ohmine,Jansen}. The semiclassical
expansion circumvents these calculations in a very profound way.
Corrections to the trajectory to low order in $\hbar$ carry the
necessary information. Combining several semiclassical
trajectories~\cite{148} allows them to interfere and the leading
order in $\hbar$ ($\hbar^{n}$ for $R^{n}$) survives and gives the
classical response. This allows us to avoid computing stability
matrices which is required when the classical limit is considered
from the outset. The classical limit obtained this way reproduces
the results of mode coupling theory and removes all ambiguities as
to how higher order correlation functions should be
factorized~\cite{vanZon,Bouch,Gotze,Keyes2}.

To illustrate how this works let us consider the following model
Hamiltonian $H_m$
~\cite{Khidekelb,Chernyakb,43,Chernyak340,Mukamel358}
\begin{equation}\label{2.1t}
  H_m = \sum_j \left (\frac {P_j^2} {2M_j} + \frac {M_j \Omega_j^2Q_j^2}
   {2} \right )  + V({\bf Q}),
\end{equation}
where $P_j(Q_j)$ is the momentum (coordinate) operator of the j'th
primary mode, $\Omega_j$ and $M_j$ are its frequency and reduced
mass respectively and $V({\bf Q})$ is the anharmonic part of the
potential. The primary modes interact with a large number of
harmonic (bath) coordinates which induce relaxation and dephasing.
Low-frequency bath degrees of freedom {\bf q} and their coupling
to the primary modes are described by the Hamiltonian $H_B$ and
the material Hamiltonian is given by~\cite{Mukamel,Mukamel358}
\begin{equation}\label{2.6t}
H = H_m ({\bf Q}) +  H_B ({\bf Q}, {\bf q}).
\end{equation}

We assume a harmonic bath linearly coupled to the primary
coordinates $Q_j$.
\begin{equation}\label{2.7t}
  H_B = \sum_{\alpha} \left [ \frac {p_{\alpha}^2} {2m_{\alpha}}
      + \frac {m_{\alpha} \omega_{\alpha}^2} {2}
      \left(q_{\alpha} - \sum_j\frac {c_{j\alpha}} {m_{\alpha} \omega^{2}_{\alpha}}
       Q_j\right)^2 \right ],
\end{equation}
where $p_{\alpha}(q_{\alpha})$ are momentum (coordinate)
operators of bath oscillators.

This model gives the following Brownian oscillator form for the
spectral density~\cite{43}
\begin{eqnarray}\label{ghi}
{C''}(\omega) & = & \textrm{Im}\left(\frac{1}{{M}
({\Omega}^2 +\omega{\Sigma} (\omega)-{I}{\omega}^2 +  i\omega\mathbf{\gamma}(\omega))}\right).
\end{eqnarray}
${M}$, ${\Omega}$ and ${E}$ are diagonal matrices and their matrix elements are $M_{ij}=\delta_{ij}M_j$, $\Omega_{ij}=\delta_{ij}\Omega_j$ and $I_{ij}=\delta_{ij}$.
$C''(\omega)$ is the odd part of the spectral density, which is related to the even part $C'(\omega)$ by the fluctuation-dissipation theorem. The spectral density (Eqs. (\ref{4.8t}) and (\ref{4.8s})) is given by $C(\omega)=[1+\coth(\beta\hbar\omega/2)]C''(\omega)$.

$\gamma_{ij}$ ($\Sigma_{ij}$) is the imaginary (real) parts of a
self energy operator representing relaxation (level shift).
\begin{eqnarray}\label{ghi1}
 \gamma_{ij} (\omega) & = &
\frac{\pi}{M_{i}}\sum_{\alpha}
\frac{c_{j\alpha}c_{i\alpha}}{2m_{\alpha}\omega^{2}_{\alpha}} [
\delta(\omega-\omega_{\alpha}) + \delta(\omega+\omega_{\alpha})],
 \nonumber\\
 \Sigma_{ij}(\omega) & = &-\frac{1}{\pi}{\text Re }\int_{-\infty}^{\infty}d\omega'
 \frac{\gamma_{ij}(\omega')}{\omega'-\omega},
\end{eqnarray}

Equations~(\ref{65}) and (\ref{68}), together with
(\ref{4.8t}), (\ref{4.8s}) and (\ref{ghi}) constitute closed expressions for the
Brownian oscillator response functions. Ordinary Langevin
equations are obtained by taking the overdamped limit
$\gamma>>\Omega$ of Eq~(\ref{ghi}).
When the primary coordinates are uncorrelated the superoperator Greens functions are
\begin{eqnarray}
G^{+-}_{ij}(\tau) & = & 2 \theta(\tau) \exp(-\Lambda_i\tau) \Lambda_i\lambda_i\delta_{ij}\\ \label{eq:langevinpp}
G^{++}_{ij}(\tau)
& = & \hbar  \exp(-\Lambda_i\tau)\Lambda_i\lambda_i\coth(\beta\hbar\Lambda_i)
\delta_{ij}
+\frac{4}{\beta}
\sum_{n=1}^{\infty}\frac{\nu_n\exp(-\nu_n \tau)}{\nu_n^2-\Lambda_i^2}\Lambda_i\lambda_i\delta_{ij}
\end{eqnarray}
where $\nu_n\equiv{2\pi n}/{\hbar\beta}$ are the Matsubara frequencies,
$\Lambda_i=\Omega_i^2/\gamma_{ii}$ and $\lambda_i=1/M_i\Omega_i^2$.
The expansion of nonlinear
response functions using collective coordinates have been
discussed in detail~\cite{Mukamel,Chernyakb,Khidekelc,Fried} and
recently employed in mode coupling theory~\cite{vanZon,Denny2}.

All nonlinear response functions of the linearly driven harmonic
oscillator vanish identically due to interference among Liouville
space paths~\cite{Mukamel}.  The simplest model that shows a
finite nonlinear response is a nonlinearly driven Harmonic
oscillator where the operator $A$ is a nonlinear function of the
coordinate. This model has been studied both quantum mechanically
and classically~\cite{Tanimura}. Its response can be alternatively
computed by following the dynamics of the Gaussian wavepackets in
the complete (system and bath) phase space, since the system-bath
Hamiltonian $H_B$ is harmonic in the full phase space
$\{P,Q,p_\alpha,q_\alpha\}$~\cite{Reichman2,vanZon,keyes,Keyes2,c}.

We next discuss the connection between our results and a fully
classical computation of the response. In classical mechanics the
density matrix assumes the form of an ordinary distribution
function in phase space. This can be obtained from the quantum
density matrix by switching to the Wigner
representation~\cite{Schleich}.
\begin{equation}
\label{Before}\rho_{\small W}({\bf p}{\bf q};t)=
\frac{1}{(2\pi\hbar)^{d}}\int d{\bf s}\langle{\bf q}-{\bf
s}/2|\rho (t) |{\bf q} +{\bf s}/2\rangle \exp(\mathit{{\bf
p}}.{\bf s})
\end{equation}
where $d$ is the number of degrees of freedom. The Wigner
representation offers a transparent and simple semiclassical
picture that interpolates between the quantum and classical
regimes. Wavefunctions, on the other hand, do not have a clear
classical counterpart (although there are, of course, very
powerful semiclassical approximations for the wavefunction such as
the WKB approximation).

Wick's theorem for superoperators in Liouville space allows to
develop a unified picture of quantum field and classical mode
coupling theories which clearly reveals the information content of
classical and quantum nonlinear response functions. Both classical
and quantum response functions contain interference. Quantum
mechanically it is between $2^n$ Liouville space paths. The
classical interference is of a very different nature~\cite{prl}and
involves $2^n$ closely lying trajectories. The response function
in phase space is obtained by an ensemble averaging over such
bundles of trajectories~\cite{Khidekelc,prl}. Alternatively, the
classical response  can  be recast using stability matrices which
carry the relevant dynamical information on the vicinity of a
given trajectory. The connection between the quantum and classical
$2^{n}$-fold interference is more transparent by keeping the
left/right or the $+/-$ pathways rather than working in phase
space. We keep $\hbar$ alive during the semiclassical calculation
and send it to zero only at the end.

In the fully classical phase space approach, we take the two
separate (left and right) paths required in a quantum mechanical
formalism and expand them around a single classical reference
path, letting the stability matrices carry the burden of retaining
the information about the differences between the paths. In the
present $+/-$ representation we keep track of closely lying
trajectories by retaining terms to order $\hbar$ and combine them
only at the  very end. This way stability matrices which pose
enormous computational difficulties~\cite{Dellago} never show up.

Another way to view the classical/quantum connection is by
starting with the expressions for quantum mechanical nonlinear
response functions in terms of combinations of $n$ point
correlation functions of the relevant variables. These correlation
functions differ by their time ordering i.e. $\langle A(\tau_1)
A(\tau_2) A(\tau_3)\rangle$, $\langle A (\tau_2) A(\tau_1)
A(\tau_3)\rangle$ etc. $R^{(n)}$ is then a combination of $2^n$
such correlation functions, each representing a distinct Liouville
space pathway. Classically, of course, time ordering is immaterial
since all operators commute and we have only a single $n$ point
correlation function. The presence (absence) of symmetry with
respect to the permutation of the $n$ time variables in classical
(quantum) correlation functions implies that the effective
multidimensional space of time is reduced by a factor of $n!$ in
the classical case. Classical correlation functions thus carry
less information than their quantum counterparts. Classically, it
suffices to calculate $\langle A(\tau_1) A(\tau_2)\cdots
A(\tau_n)\rangle$ for $\tau_1 \le \tau_2 \cdots \le \tau_n$.
Quantum mechanically, each of the $n!$ permutations of the time
arguments is distinct and carries additional information. The
stability matrices provide the extra information required for
computing the response functions from classical correlation
functions.

Since classical correlation functions do not carry enough
information for computing nonlinear response functions, it is not
possible to simulate and interpret the response in terms of
standard equilibrium fluctuations; additional nonequilibrium
information~\cite{Martens} is necessary.

Correlation functions are equilibrium objects and can be computed
using sums over unperturbed trajectories; response functions can
be obtained either as equilibrium averages (stability matrices) or
recast in terms of $2^n$ closely lying nonequilibrium trajectories
perturbed at various points in time. Quantum corrections to
classical response functions may be represented in terms of
classical response functions of higher orders~\cite{Khidekelc}.

Finally we note that an alternative semiclassical ${\hbar}$
expansion of the response is possible even when the temperature is
low compared with the material frequencies, and the system is
highly quantum, provided anharmonicities are
weak~\cite{Chernyaka,Chernyak340}. The leading terms in the
expansion can be obtained by solving classical equations of
motion. This is done by  hiding the $\hbar$ in the $\coth$ factor
in Eq.~(\ref{4.8t}) by recasting it in the form
$\coth(\omega/2\omega_{T})$ where $\omega_{T}=k_{B}T/\hbar$ is the
thermal frequency, and redefining $G^{++}$ by multiplying it by
$\hbar$. The response is then analytic in $\hbar$ (as long as we
forget about the $\hbar$ dependence of $\omega_{T}$).
Semiclassical approximations ordinarily hold when the temperature
is high compared to all relevant vibrational frequencies. This
points to a much less obvious, low-temperature weak anharmonicity
regime, where the response of the system behaves almost
classically even though its temperature is very low.

\section{Wick's theorem for Fermion Superoperators}

One reason why the handling of Boson operators and fields is
simpler compared to Fermions is that superoperators corresponding
to elementary Boson operators are also elementary i.e., their
commutators are also numbers. To see that let us consider the
commutation rules of superoperators by acting with commutators on
an arbitrary operator $F$.
\begin{eqnarray}
[Q_{jL},Q_{kL}] F &=& [Q_{j},Q_{k}]F\\\nonumber
[Q_{jR},Q_{kR}]F&=&-F [Q_{j},Q_{k}]\\\nonumber [Q_{jL},Q_{kR}] &=&
0
\end{eqnarray}
Since $[Q_{j},Q_{k}]$ is a number, we see that the corresponding
superoperators are elementary as well. This property holds also if
we consider the linear combinations in the $+/-$ representation.
To see that we start with $[A_{+}, B_{-}]$ and act on an ordinary
operator $F$
\begin{eqnarray}
\label{78}
 [A_{+}, B_{-}]F &=& \frac{1}{2}[A,B] F + \frac{1}{2} F [A,B]\\\nonumber
[A_{-}, B_{-}]F &=& 4 [A_{+},B_{+}] F =  [[A,B], F].
\end{eqnarray}
Since the commutator of elementary operators is a number,
Eq.~(\ref{78}) gives
\begin{eqnarray}
[A_{+}, B_{-}] &=&  [A,B]\\\nonumber [A_{-}, B_{-}]&=& 4 [A_{+},
B_{+}]=0
\end{eqnarray}

The commutators of elementary Boson superoperators are thus
numbers. It then follows that the superoperators corresponding to
elementary Boson operators are Gaussian, be it in the $+/-$ or the
$L,R$ representation. The indices $\nu$ in Eq.~(\ref{69}) can thus
run over $+/-$ or $L,R$ and Wick's theorem holds in either case.

Using $L,R$, the functional can be used to generate individual
Liouville pathways. Using $+-$  it generates combinations of such
paths, making the classical limit more transparent, since we work
with combinations of Hilbert space correlation functions which
enter the response and have well defined classical limits.

Life is more complicated for Fermions. To see that let us consider
the anti-commutation rules for Fermi elementary superoperators.
\begin{eqnarray}
\{Q_{jL},Q_{kL}\} F &=& \{Q_{j},Q_{k}\}F\\\nonumber
\{Q_{jR},Q_{kR}\} F &=& F\{Q_{j},Q_{k}\}\\\nonumber
\{Q_{jL},Q_{kR}\} F &=& 2 Q_{j} FQ_{k}
\end{eqnarray}
Since $\{Q_{j},Q_{k}\}$ is a number this shows that left/right or
right/left superoperators have elementary anticommutators but the
anticommutator of left/right operators is not generally a number.
We thus do not have a Gaussian statistics. Note however that left
and right operators always commute. The generating functional
needs to be formulated using Grassman algebra of anticommuting
numbers similar to standard Green functions~\cite{Negele,Mahan}.
The important point is that a modified Wick theorem still holds
for Fermi superoperators. It is given by Eq.~(\ref{69}) with the following changes:\\
(i) We must use $L,R$ rather than $+/-$ variables for $\nu$.\\
(ii) Each term needs to be multiplied by $(-1)^{P}$ where $P$ is
the number of permutations of elementary operators required to
bring them to the specified order. Since left and right
superoperators commute, we only count the number of permutations
among left and among right operators. (Permuting a right and left
operator does not count in $P$).

The expectation  of the $T$ product of any number of (Boson or
Fermion) superoperators may thus be expressed as the sum of all
possible products of expectations of $T$ products of the separated
pairs of operators for the reference many-particle density matrix
$\rho _{0}$ corresponding to $H_{0}$.

\section{Discussion}

Hilbert space and Liouville space offer a very different language
for the description of nonlinear response. The density matrix
provides a fully time ordered description since we only need to
propagate it forward in time. In contrast, the wavefunction
involves both forward and backward propagations. The choice is
between following the ket only, moving forward and backward or
following the joint dynamics of the ket and the bra and moving
only forward. Artificial time variables (Keldysh loops) commonly
used in many-body theory~\cite{Matsubara,Keldysh,Schwinger} are
connected with the wavefunction picture. The density matrix uses
the real laboratory timescale throughout the calculation.

In Liouville space all observables are time-ordered leading
naturally to a semiclassical approximations~\cite{Sepulveda} and
Feynman path integral diagrammatic
techniques~\cite{Negele,Fetter2,Doniach,Abrikosov2,Kadanoff,Mahan,Haug}.
Maintaining time ordering allows to recast $S^{(n)}$ using
nonlinear response functions which decouple the field and the
material parts. The nonlinear response function is calculated as a
path integral in Liouville space by summing over the various
possible pathways in Liouville-space which contribute to the
polarization.  Path integrals have been extensively used as a
useful tool for numerical computations of mixed quantum-classical
calculations~\cite{Schulman,feynman}.  The density matrix
formulation provides a similar development for phase space-based
numerical procedures.  Graphic visualization of these paths is
provided by double-sided Feynman diagrams~\cite{Mukamel}.

The density matrix Liouville Space picture offers many attractive
features. Physical observables are directly and linearly related
to the density matrix.  Consequently, every step and intermediate
quantity appearing in the description has a simple physical
meaning and a clear classical analogue. This should be contrasted
with wavefunction-based calculations of a transition amplitude,
which by itself is not an observable since signals are related to
sums of products of such amplitudes.

The density matrix provides a practical way for performing
ensemble averagings and developing reduced descriptions where bath
degrees of freedom are traced out from the outset. Since it
represents the state of the system by a matrix rather than a
vector, an $N$ point grid for ${\bf p}$ and ${\bf q}$ in a
semiclassical picture will require $N^2$ points for the
wavefunction and $N^4$ for the density matrix.  The ability to
perform ensemble averagings and reduced descriptions more than
compensates for this price for complex systems. Many-body theory
of superoperators further naturally allows for the treatment of
dephasing and decoherence effects. Diagonal and off diagonal
elements of the density matrix are known as {\em populations} and
{\em coherences}, respectively.  When an off diagonal element
evolves in time for a system coupled to a bath, it acquires a
phase since its evolution from the left (ket) and the right (bra)
is governed by different bath Hamiltonians.  This phase depends on
the state of the bath.  When we perform an ensemble average of
these elements over the distribution of the bath degrees of
freedom, this variable phase results in a damping of these
elements.  The damping of off diagonal elements of the density
matrix resulting from phase (as opposed to amplitude) fluctuations
is called {\em pure dephasing} or phase relaxation (also known as
{\em decoherence}). Dephasing processes can only be visualized in
Liouville space by following simultaneously the evolution of the
bra and of the ket and maintaining the bookkeeping of their joint
state. Dephasing processes directly affect all spectroscopic
observables since they control the coherence which is the window
through which the system is observed. Different pathways
representing distinct sequences of populations and coherences, are
naturally separated in Liouville space.

\section*{Acknowledgements}

The support of the National Science Foundation grant no.
CHE-0132571 is gratefully acknowledged. I wish to thank Dr.
Vladimir Chernyak for most useful discussions and Dr. Thomas la
Cour Jansen and Adam Cohen for helpful comments.

\appendix

\section{Some Useful Relations for Superoperator Algebra in Liouville Space}

A Liouville space operator $A_\alpha$ is labelled by a Greek
subscript where $\alpha =L, R, +, -$. It is defined by its action
on $X$, an ordinary (Hilbert space) operator. We write a general
matrix element
\begin{equation}
\label{A1} (A_\alpha X)_{ij} \equiv \sum_{\kappa\ell}
(A_\alpha)_{i j, \kappa \ell} X_{\kappa\ell}
\end{equation}
$A_\alpha$ is thus a \emph{tetradic} operator with four indices.
Since $A_L X\equiv A X$ and $A_R X \equiv X A$ we obtain using
Eq.~(\ref{A1})
\begin{equation}
\label{A2} (A_L)_{ij, \kappa\ell} = A_{i\kappa} \delta_{j\ell}
\end{equation}
\begin{equation}
\label{A3}
 (A_R)_{ij, \kappa\ell} = A_{\ell j} \delta_{i\kappa}
\end{equation}
Note that the order of the $j\ell$ indices in Eq.~(\ref{A3}) has
been reversed.

Since $A_+ \equiv \frac{1}{2}(A_L + A_R)$ and $A_-\equiv
A_L-A_R\,$ we have $(A_{-})_{ij,\kappa\ell} = A_{i\kappa}
\delta_{j\ell} - A_{\ell j} \delta_{i\kappa}$ and
$(A_{+})_{ij,\kappa\ell} = \frac{1}{2}[A_{i\kappa} \delta_{j\ell}
+ A_{\ell j} \delta_{i\kappa}]$. It then follows that $[A_L,
B_R]=0$. This commutativity of left and right operators is
possible thanks to the large size of Liouville space and
simplifies algebraic manipulations resulting in many useful
relations.
\begin{eqnarray}
& 2[A_+, B_-]=[A_L, B_L] - [A_R, B_R]= (AB)_- -(BA)_-,\\\nonumber
& \exp (A_L) =\exp (A_+) \exp(\frac{1}{2}A_-),\\\nonumber & \exp
(A_+) = \exp (\frac{1}{2}A_L) \exp (\frac{1}{2}A_R),\\\nonumber &
\exp(A_-)=\exp(A_L) \exp (-A_R),\\\nonumber & (\exp A)_{+} = 2\exp
(A _{+}) \cosh (\frac{1}{2}A _{-}),\\\nonumber & (\exp A)_{-} =
2\exp (A _{+}) \sinh (\frac{1}{2}A _{-}).
\end{eqnarray}

In the following $a$ is a complex number
\begin{eqnarray}
\delta (\omega-A_-)&=&\int da \,\,\delta (a-A_L) \delta (\omega-a
+ A_R),\\\nonumber \delta (\omega-A_+)&=&\int da \,\,\delta
(a-A_L) \delta (\omega-a-A_R),\\\nonumber  \delta(a-A_L)
\delta(a-A_R)&=&\delta(A_+-a) \delta(A_-).
\end{eqnarray}


\end{document}